# Estimating Permeability of Carbonate Rocks from NMR $T_2$ Distribution and Formation Factor Using a Physics-Based Model

Behzad Ghanbarian[1,2,3]*, Silvia L. B. Bermudez[4,5], Andre Souza[4], Gabriel S. S. Ribeiro[5], Maira C. O. Lima[5], and Paulo Couto[5]

[1] iResearchE[3] Lab, Department of Earth and Environmental Sciences, University of Texas at Arlington, Arlington TX 76019 USA

[2] Department of Civil Engineering, University of Texas at Arlington, Arlington TX 76019, United States

[3] Division of Data Science, College of Science, University of Texas at Arlington, Arlington TX 76019, United States

[4]Coordenação de Geofísica, Observatório Nacional, Ministério de Ciência, Tecnologia e Inovação, 20921-400, Rio de Janeiro, RJ, Brazil.

[5]Department of Civil Engineering/COPPE, Federal University of Rio de Janeiro, Rio de Janeiro RJ, Brazil

[*] Corresponding author's email address: ghanbarianb@uta.edu

**Abstract**

Nuclear magnetic resonance (NMR) logging provides a non-destructive method for characterizing pore structure and fluid distribution in reservoir rocks. Conventional permeability models, such as the Schlumberger-Doll Research (SDR) and Timur–Coates (TC) equations, have been widely used to estimate permeability from NMR $T_2$ relaxation time distributions. However, these empirical models require calibration against core-measured permeability, and their fitting coefficients often vary substantially among different lithologies and pore systems. In this study, we formulate and evaluate an NMR-informed implementation of a physically-based permeability relationship, i.e. critical path analysis (CPA), by linking the characteristic pore length scale to the dominant NMR relaxation time. We then estimated permeability ($\boldsymbol{k}$) from NMR $T_2$ distribution and formation resistivity factor ($\boldsymbol{F}$) for 14 carbonate rock samples collected from the Morro do Chaves Formation in the Sergipe-Alagoas Basin, Brazil. Its predictive performance was compared against the calibrated SDR and Timur–Coates models. The proposed CPA–NMR formulation achieved an RMSLE of 0.56 for 14 carbonate samples without calibration against measured permeability from the investigated formation. These findings demonstrate the potential of combining NMR relaxation measurements with electrical formation factor for physics-based permeability estimation, although further evaluation using independent datasets is required to establish applicability across different formations and lithologies.

## 1. Introduction

Permeability ($k$) estimation is fundamental to aquifer and reservoir characterization because it directly governs fluid flow and transport in geologic formations. In carbonate reservoirs, permeability prediction is particularly challenging due to complex and frequently vuggy pore structures, multiscale porosity systems, and diverse diagenetic alterations (Pirrone and Galli, 2018; Wang et al., 2018). Although conventional core-based measurements provide reliable permeability values, they are expensive, time-consuming, and spatially limited. Consequently, numerous predictive relationships have been developed to estimate carbonate permeability from more accessible rock properties, including porosity and specific surface area (Cardona and Santamarina, 2020), porosity and lithology (Male and Duncan, 2020), and mercury intrusion capillary pressure (MICP) measurements (Hulea and Nicholls, 2012; Rashid et al., 2015). However, although porosity and lithology information may often be obtained from well logs, in situ determination of specific surface area and capillary pressure characteristics remains difficult.

Nuclear magnetic resonance (NMR) measurements provide a non-destructive alternative for pore-scale characterization that can be performed both in laboratory experiments and through downhole logging tools. NMR-derived transverse relaxation-time ($T_2$) distributions are commonly interpreted as indicators of pore-size distribution and fluid distribution within the pore network, thereby providing a basis for permeability estimation (Kenyon et al., 1988; Coates et al., 1999). For a fully saturated porous medium, the measured NMR signal arises primarily from hydrogen-bearing fluids occupying the pore space, such that the signal amplitude associated with a given relaxation component is related to the volume of fluid contributing to that component. Under fast-diffusion conditions and when surface relaxation dominates over bulk and diffusion-induced relaxation, the transverse relaxation rate can be approximated as $1/T_2 \approx \rho_2 S/V$, where $\rho_2$ is the transverse surface relaxivity and (S/V) is the pore surface-to-volume ratio (Coates et al., 1999). Consequently, larger pores generally exhibit longer $T_2$ values, whereas smaller pores relax more rapidly. The measured $T_2$ distribution can therefore provide a proxy for the distribution of pore-scale surface-to-volume ratios, provided that surface relaxivity does not vary strongly within the sample and that the assumptions underlying the fast-diffusion approximation are reasonably satisfied. Importantly, however, an NMR-derived pore-size distribution does not necessarily represent the pore-throat size distribution controlling permeability, particularly in heterogeneous carbonate rocks.

One of the most widely used NMR-based permeability models is the Timur–Coates (TC) equation, which relates permeability to porosity ($\phi$), free fluid index (FFI), and bulk volume irreducible (BVI) through the empirical power-law relationship (Coates et al., 1999)

$$k = C_{TC}\phi^4\left(\frac{FFI}{BVI}\right)^2 \tag{1}$$

where $C_{TC}$ is an empirical proportionality coefficient. The free fluid index (FFI) represents the fraction of NMR-detectable pore fluid associated with relaxation times above a prescribed $T_2$ cutoff and is commonly interpreted as relatively mobile fluid, whereas bulk volume irreducible (BVI) represents the fluid fraction below the cutoff and is associated with more strongly confined or capillary-bound fluid. The TC model attempts to distinguish between mobile and immobile fluids within the pore system, but its formulation remains fundamentally empirical.

Another widely applied NMR-based permeability relationship is the Schlumberger-Doll Research (SDR) model, which relates $k$ to $\phi$ and the logarithmic mean relaxation time ($T_{2lm}$) as follows (Kenyon et al., 1988):

$$k = C_{SDR}\phi^4 T_{2lm}^2 \quad (2)$$

where $C_{SDR}$ is an empirical constant whose value depends strongly on lithology, pore geometry, and measurement conditions.

Behroozmand et al. (2015) reviewed the application of NMR methods to near-surface characterization and highlighted that the predictive reliability of SDR-based permeability estimation depends on lithology and calibration strategy. Their analysis showed that the SDR model can provide reliable predictions for inorganic unconsolidated sediments when properly calibrated at site- or geology-specific scales, whereas it fails in organic-rich materials such as peat, where relaxation time no longer correlates with pore-size-controlled flow. Their study further demonstrated that generalized calibration across mixed lithologies is ineffective and that simplified formulations excluding porosity may reduce uncertainty under conditions where porosity estimation is biased by magnetic susceptibility effects.

Both the TC and SDR models, Eqs. (1) and (2), rely heavily on calibration against core-measured permeability data. Although these models have been successfully applied to sandstone reservoirs (Coates et al., 1999; Chen et al., 2023), their applicability becomes considerably more challenging in formations characterized by heterogeneous pore structures and complex mineralogy, particularly in carbonate reservoirs. This limitation arises because the coefficients $C_{CT}$ and $C_{SDR}$ depend strongly on pore geometry, mineralogical composition, and surface relaxivity (Mohnke and Yaramanci, 2008), thereby restricting their transferability across different formations.

To improve predictive flexibility, both models have been generalized by replacing the fixed exponents in Eqs. (1) and (2) with adjustable fitting parameters calibrated separately for each formation or dataset (Ren et al., 2019; Gao et al., 2021; Weller and Zhang, 2022). Table 1 summarizes several generalized TC and SDR formulations reported for carbonate reservoirs worldwide. As shown, the calibrated exponents vary substantially among formations, further highlighting the strong lithology dependence of empirical NMR-based permeability relationships.

More recently, Mashhadi et al. (2025) evaluated SDR-based permeability estimation for unconsolidated geological materials using 92 laboratory samples and 198 borehole screens collected from sites in Denmark and Germany. Their dataset included inorganic sediments (sand, clay, and till) as well as organic materials such as peat and gyttja. They found that the SDR model performed reasonably well for inorganic materials when geology-specific calibration was applied, whereas it failed for peat soils, where no meaningful correlation existed between relaxation time and permeability. Although gyttja followed SDR-type scaling behavior, it required calibration coefficients substantially different from those of inorganic sediments. Their results further demonstrated that generalized calibration across mixed lithologies was ineffective, whereas site- or geology-specific calibration significantly improved prediction accuracy.

Currently, the exponents in generalized TC and SDR formulations are commonly optimized to maintain an approximately constant prefactor at the formation scale.

However, Ghanbarian et al. (2027) recently argued that such treatment may lead to physically inconsistent interpretations because variations in pore geometry, surface relaxivity, and electrical connectivity are implicitly absorbed into empirical exponents.

Previous studies have established theoretical connections between NMR relaxation measurements and permeability through characteristic pore dimensions and pore-scale transport relationships. In particular, Sigal (2002) developed an NMR-informed permeability expression motivated by percolation theory (see his Eq. (8)), relating permeability to a characteristic NMR relaxation time and porosity through Archie's relationship (Archie, 1942). Although this formulation provided a theoretical interpretation of SDR-type permeability models, its predictive performance was not directly evaluated using measured permeability data. Moreover, replacing formation factor with Archie's relationship introduces an additional assumption that may not adequately describe electrically complex carbonate rocks. Building on this theoretical foundation, the present study integrates critical path analysis with the dominant NMR transverse relaxation time, effective surface relaxivity, and independently measured formation factor. The resulting formulation retains an explicit electrical transport constraint without requiring Archie's relationship and is quantitatively evaluated against measured permeability for 14 heterogeneous carbonate samples. The contribution of this study is therefore the implementation and evaluation of a physically motivated NMR–CPA framework, rather than the derivation of an entirely new permeability law.

The main objective of this study is to develop and evaluate a physics-based CPA–NMR formulation for permeability estimation that does not require calibration against measured permeability from the target formation. Specifically, we investigate whether the dominant NMR transverse relaxation time, combined with effective surface relaxivity and independently measured electrical formation factor, can provide a practical approximation of the characteristic length scale required by CPA. The predictive performance of the proposed formulation is evaluated using 14 heterogeneous carbonate samples from the Morro do Chaves Formation and compared with existing SDR and Timur–Coates permeability models.

Table 1. The generalized TC and SDR models and their parameters calibrated for carbonate rocks from various reservoirs. $k$ is the permeability, $\phi$ is the porosity, $FFI$ represents the free fluid index, $BVI$ denotes the bulk volume irreducible, and $\rho_2$ is the surface relaxivity. The unit of $\rho_2$ in the Hassall et al. (2004) model is $\mu m/s$.

| Reference | $k$ unit | $\phi$ unit | $T_{2lm}$ unit | Model | Reservoir |
|---|---|---|---|---|---|
| Amabeoku et al. (2001) | mD | $cm^3cm^{-3}$ | ms | $k = 0.0574\phi^{3.79}T_{2lm}^{1.89}$ | Arab carbonate reservoir |
| Amabeoku et al. (2001) | mD | $cm^3cm^{-3}$ | ms | $k = 511.4\phi^{2.84}\left(\frac{FFI}{BVI}\right)^2$ | Arab carbonate reservoir |
| Hassall et al. (2004) | mD | $cm^3cm^{-3}$ | ms | $k = 0.5\phi^2(\rho_2 T_{2lm})^2$ | Abu Dhabi, UAE |
| Machado et al. (2008) | mD | % | ms | $k = 7.3 \times 10^{-11}\phi^{2.68}T_{2lm}^{4.56}$ | Santos Basin, Brazil |
| Machado et al. (2008) | mD | % | ms | $k = \left(\frac{\phi}{109.64}\right)^{1.74}\left(\frac{FFI}{BVI}\right)^{3.85}$ | Santos Basin, Brazil |
| Agha et al. (2018) | mD | $cm^3cm^{-3}$ | ms | $k = 0.0216\phi^{1.64}T_{2lm}^{1.68}$ | Asmari, Iran |
| Agha et al. (2018) | mD | $cm^3cm^{-3}$ | ms | $k = \left(\frac{\phi}{0.222}\right)^{3.9}\left(\frac{FFI}{BVI}\right)^{0.51}$ | Asmari, Iran |
| Martins (2019) | mD | % | ms | $k = 0.12 \times 10^4\phi^{4.48}T_{2lm}^{1.0}$ | Morro do Chaves, Brazil |
| Martins (2019) | mD | % | ms | $k = 3.45 \times 10^4\phi^{5.05}\left(\frac{FFI}{BVI}\right)^{0.78}$ | Morro do Chaves, Brazil |
| Shabani et al. (2021) | mD | % | ms | $k = 0.0004\left(\frac{\phi}{10}\right)^3 T_{2lm}^{1.6}$ | Asmari, Iran |
| Shabani et al. (2021) | mD | % | ms | $k = 0.022\left(\frac{\phi}{10}\right)^4\left(\frac{FFI}{BVI}\right)^2$ | Asmari, Iran |

## 2. NMR-Informed Permeability Framework

Unlike empirical permeability formulations, physics-based models attempt to explicitly relate permeability to pore geometry and transport properties of the porous medium (David et al., 1990; Bernabé et al., 2011; Ghanbarian and Javadpour, 2017). Within the framework of critical path analysis (CPA), Katz and Thompson (1986) developed a percolation-based permeability relationship involving a critical pore diameter ($d_c$) and formation factor ($F$) as follows

$$k = \frac{d_c^2}{C_{CPA} F} \quad (3)$$

where $C_{CPA}$ is a proportionality constant. In Eq. (3), $d_c$ represents the characteristic constriction size associated with the onset of a sample-spanning connected pathway, rather than an arithmetic or volumetric mean pore diameter. Katz and Thompson (1986) originally proposed a value of 226 for $C_{CPA}$; however, subsequent studies reported different values depending on pore structure and porous medium type. A review of reported $C_{CPA}$ values can be found in Ghanbarian et al. (2016). Eq. (3) has been successfully applied to estimate the $k$ of sandstone and carbonate rocks from mercury intrusion porosimetry data (Katz and Thompson, 1986; Ghanbarian et al., 2016; Nishiyama and Yokoyama, 2017; Meng et al., 2026).

Katz and Thompson (1986) further suggested that the critical pore diameter may be estimated from the inflection point of the capillary pressure curve, corresponding to the mode of the pore-throat size distribution (Ghanbarian and Skaggs, 2022). In the present study, we set

$$d_c = c_s \rho_{2e} T_{2peak} \quad (4)$$

in which $c_s$ is a pore-geometry factor equal to 4 for cylindrical pores and 6 for spherical pores, $\rho_{2e}$ is the effective surface relaxivity (Coates et al., 1999), and $T_{2peak}$ represents the relaxation time corresponding to the maximum incremental porosity of the NMR $T_2$ distribution.

The approximation $d_c = c_s\rho_{2e}T_{2peak}$ requires careful interpretation because NMR transverse relaxation is primarily sensitive to pore surface-to-volume characteristics, whereas the critical diameter, $d_c$, in CPA represents a flow-controlling pore-throat scale along the sample-spanning flow network. As discussed by Coates et al. (1999), NMR-derived characteristic pore dimensions do not necessarily correspond to critical pore-throat dimensions, particularly in heterogeneous carbonate rocks with complex pore-body-to-throat relationships. To account phenomenologically for this distinction, an effective surface relaxivity, $\rho_{2e}$, is introduced to relate the NMR-derived characteristic length scale to the hydraulically relevant pore-throat dimension. Unlike the intrinsic transverse surface relaxivity, $\rho_2$, which characterizes relaxation at the pore-fluid–solid interface, $\rho_{2e}$ represents an effective conversion parameter that incorporates the relationship between NMR-sensitive pore dimensions and flow-controlling constrictions. Accordingly, Eq. (4) constitutes a modeling approximation rather than a direct measurement of the critical pore-throat diameter, and its validity depends on the correspondence between NMR relaxation characteristics and the pore-network geometry governing fluid flow.

Substituting Eq. (4) into Eq. (3) yields

$$k = \frac{c_s^2\rho_{2e}^2}{C_{CPA}}\frac{T_{2peak}^2}{F} \tag{5}$$

Eq. (5) provides a physically motivated relationship linking permeability to pore geometry, electrical connectivity, and NMR relaxation behavior. The use of $T_{2peak}$ as a characteristic relaxation parameter has been previously explored in several studies. For example, Chi and Roth (2018), Lai et al. (2018), and Weller and Zhang (2023) successfully employed $T_{2peak}$ in permeability-related analyses. Rezaee et al. (2012) further demonstrated that permeability in tight gas sandstones was strongly correlated with porosity and $T_{2peak}$ using multiple linear regression analysis.

Although CPA-based permeability relationships have been extensively evaluated using critical pore-throat sizes determined from MICP measurements, such measurements are not routinely available in many subsurface applications, particularly well logging. The objective of the proposed formulation is therefore to approximate the critical pore-throat length scale required by CPA using the dominant NMR relaxation time and effective transverse surface relaxivity, thereby enabling permeability estimation without direct MICP or micro-CT characterization. The effective surface relaxivity provides a phenomenological means of accounting for differences between NMR-sensitive pore dimensions and permeability-controlling pore throats. Although this approximation does not establish a unique correspondence between $T_{2peak}$ and the critical pore-throat diameter, its practical utility can be assessed through comparison of predicted and independently measured permeability. Accordingly, the present study evaluates the predictive performance of this NMR–CPA formulation rather than independently validating the underlying pore-throat length-scale correspondence.

## 3. Experiments

A total of 14 carbonate rock samples collected from the Morro do Chaves Formation in the Sergipe-Alagoas Basin, northeastern Brazil, were analyzed in this study. The Lower Cretaceous Morro do Chaves Formation consists predominantly of limestones characterized by variable porosity, heterogeneous pore systems, and locally developed vuggy structures.

Prior to petrophysical characterization, all samples were cleaned to remove residual hydrocarbons, salts, and other pore-filling contaminants. Cleaning was performed using Soxhlet extraction, with toluene to dissolve hydrocarbons and methanol to remove salts and residual water from the pore system. This procedure enabled continuous extraction while minimizing alteration of the rock framework. Following extraction, the samples were dried in an oven at 60°C for 12 h to ensure complete removal of residual fluids and solvents before measurements.

Porosity was determined using helium gas expansion based on Boyle's law. Absolute permeability was measured using steady-state gas permeametry during routine core analysis. In this method, an inert gas was injected through the core plug at a controlled pressure gradient until steady-state flow was achieved. Permeability was subsequently calculated using Darcy's law, and Klinkenberg corrections were applied to estimate equivalent liquid permeability.

The samples were then prepared for NMR measurements by full brine saturation with a 30,000 ppm KCl solution at a density of 1.067 g $cm^{-3}$. The KCl brine was selected to minimize clay swelling and fluid–rock interactions. Because the samples covered a broad permeability range, different saturation procedures were applied for high- and low-permeability samples. For samples with permeability values greater than 100 mD, saturation was achieved by applying vacuum while the samples were immersed in brine. For the lower-permeability samples (($k<100$) mD), an initial vacuum treatment was applied for 8 h, after which the plugs were transferred to a pressure cell completely filled with the saturating brine. The samples were maintained under an applied fluid pressure of up to 2,000 psi for approximately 24 h to promote complete saturation. This pressure was applied during the saturation procedure and should not be interpreted as an independently imposed external confining pressure. Electrical measurements were subsequently conducted under ambient laboratory conditions. After saturation, the samples were reweighed, and the difference between dry and saturated masses was used to estimate the volume of brine occupying the pore system. Prior to NMR measurements, the samples were wrapped with Teflon tape to preserve saturation conditions and minimize evaporative losses during acquisition.

The electrical formation factor, $F$, was determined from resistance measurements performed on fully brine-saturated core plugs using a VINCI Technologies EPS-A (Electrical Properties System at Ambient Conditions) equipped with a two-electrode configuration. The electrodes were positioned at opposite end faces of each cylindrical plug, with pneumatic loading applied to maintain reproducible electrode-to-core contact under ambient laboratory conditions without imposing external confining pressure. Electrical resistance was measured at 10 kHz using an external LCR meter, with three measurements obtained for each sample. This frequency is consistent with previous investigations of rock electrical properties. For example, Zisser and Nover (2009) reported negligible electrode polarization effects at 10 kHz in their experimental configuration, while Revil et al. (2017) obtained comparable formation factors over frequencies ranging from 10 Hz to 10 kHz (see their Figs. 9 and 10). The arithmetic mean of the three recorded resistance values, $\bar{R}$, was used to calculate the resistivity of the fully saturated rock as $R_o = \bar{R}(A/L)$, where $A$ and $L$ represent the cross-sectional area and length of the core plug, respectively. The formation factor was subsequently calculated as $F = R_o/R_w$, using the reported resistivity of the 30,000-ppm KCl saturating brine, $R_w$ = 0.147 Ω·m, at 23°C. Although the measurements were conducted under ambient laboratory conditions, the exact sample temperature and the equivalent-circuit configuration of the LCR meter were not documented. Consequently, it could not be established whether the recorded resistance represented the real component of complex impedance or another instrument-defined resistance quantity.

The NMR measurements were performed using a low-field GeoSpec2 spectrometer (Oxford Instruments, UK) operating at a magnetic field strength of 0.047 T and a resonance frequency of 2 MHz. All measurements were conducted in the NMR and Petrophysics Applications Laboratory (UFFLar), and acquisition control was implemented using GIT software (Green Imaging Technologies, Fredericton, Canada).

The transverse relaxation-time ($T_2$) measurements were acquired using the Carr–Purcell–Meiboom–Gill (CPMG) pulse sequence, which is widely used in petrophysical applications because of its sensitivity to pore-scale relaxation processes. The sequence consisted of an initial 90° radio-frequency pulse followed by a train of 180° refocusing pulses. Echo signals were recorded with signal-to-noise ratios greater than 100 to ensure reliable reconstruction of the $T_2$ relaxation-time distributions. Additional acquisition parameters comprised 16,250 echoes, an echo time of 200 μs, 64 averages, a waiting time of 10 s between averages, and a measurement temperature of 25 °C.

The measured CPMG echo-decay signals were represented as multiexponential relaxation functions, and the corresponding $T_2$ distributions were obtained through mathematical inversion using an inverse Laplace transform. The resulting distributions were represented on a common logarithmically spaced grid comprising 121 relaxation-time bins ranging from 0.01 to 10,000 ms, with a spacing of $\Delta log_{10}(T_2)$ = 0.05. The specific regularization method and associated parameter settings employed during the original NMR inversion could not be recovered from the available experimental records. The dominant relaxation time, $T_{2peak}$, was defined as the relaxation time corresponding to the maximum incremental-porosity amplitude in the $T_2$ distribution. For multimodal distributions, the global maximum was selected, irrespective of the presence of secondary peaks. This criterion was applied consistently to all 14 samples. Figure 1 presents the resulting NMR $T_2$ distributions for the investigated carbonate samples.

The bulk volume irreducible (BVI) and free fluid index (FFI) were determined from the $T_2$ distributions using a fixed $T_2$ cutoff of 100 ms for all samples, consistent with the original experimental procedure reported by Martins (2019) and Lima (2020). The integrated signal amplitude below the cutoff was assigned to BVI, whereas the integrated amplitude above the cutoff was assigned to FFI. $T_2$ cutoff values are ideally established from laboratory calibration and depend on lithology. Whereas a value near 33 ms is commonly adopted for sandstones, substantially higher and more variable cutoffs, typically within approximately 90–200 ms, have been reported for carbonate rocks (Schön, 2015). Table 2 summarizes the experimentally measured petrophysical properties of the analyzed samples, including gas porosity, NMR porosity, gas permeability, Klinkenberg-corrected permeability, and formation factor.

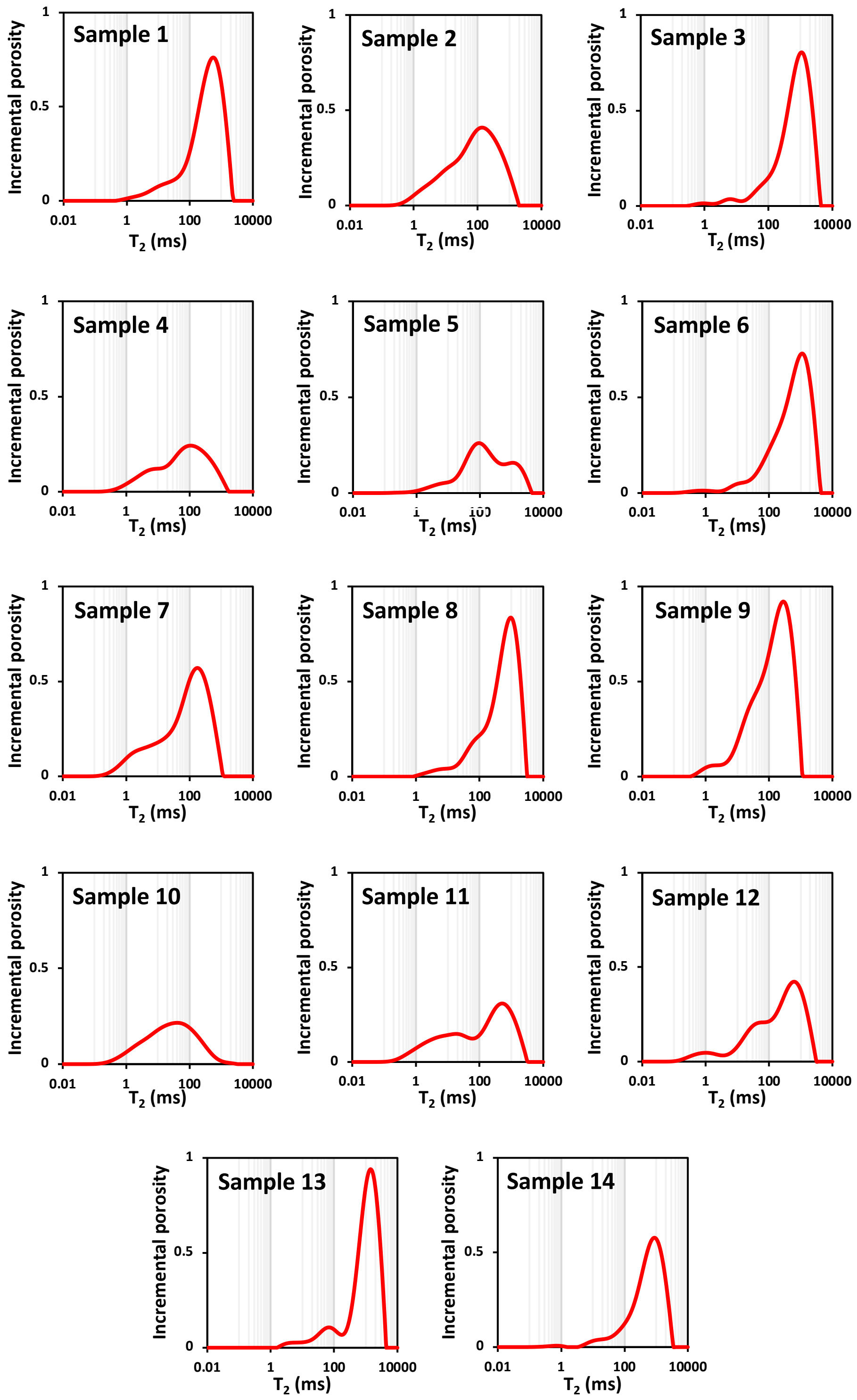

Fig. 1. The NMR $T_2$ distributions for 14 samples analyzed in this study.

Table 2. Salient properties of the carbonate samples analyzed in this study.

| Sample | $\phi_{gas}$ (%) | $\phi_{NMR}$ (%) | $k_{gas}$ (mD) | $k_{gas-Klinkenberg}$ (mD) | $F$ | $T_{2lm}$ (ms) | $T_{2peak}$ (ms) | FFI | BVI |
|---|---|---|---|---|---|---|---|---|---|
| 1 | 18.2 | 17.7 | 70.8 | 68.3 | 39.3 | 262.5 | 562.3 | 14.2 | 3.5 |
| 2 | 16.3 | 15.4 | 24.1 | 22.9 | 58.8 | 64.8 | 141.3 | 7.0 | 8.5 |
| 3 | 19.6 | 17.5 | 504.4 | 482.3 | 37.8 | 529.0 | 1122.0 | 15.5 | 2.0 |
| 4 | 10.2 | 9.4 | 5.3 | 4.7 | 57.2 | 51.0 | 112.2 | 3.8 | 5.7 |
| 5 | 8.7 | 9 | 5.2 | 4.8 | 115.7 | 133.0 | 89.1 | 5.0 | 4.0 |
| 6 | 19.8 | 17.9 | 232.9 | 227.0 | 35.1 | 437.6 | 1122.0 | 15.2 | 2.7 |
| 7 | 18.5 | 17.9 | 103.5 | 99.0 | 27.9 | 56.1 | 177.8 | 8.3 | 9.6 |
| 8 | 18.4 | 18 | 649.9 | 636.6 | 48.1 | 398.1 | 1000.0 | 15.0 | 3.0 |
| 9 | 24.9 | 24.6 | 84.0 | 81.2 | 18.9 | 105.7 | 281.8 | 14.3 | 10.3 |
| 10 | 8.7 | 8.9 | 0.4 | 0.4 | 130.2 | 23.3 | 39.8 | 1.9 | 7.0 |
| 11 | 12.2 | 12.1 | 5.7 | 4.8 | 46.1 | 75.4 | 501.2 | 6.4 | 5.7 |
| 12 | 12.1 | 13.6 | 22.1 | 18.9 | 62.3 | 145.1 | 631.0 | 8.7 | 4.9 |
| 13 | 16.8 | 16.3 | 245.1 | 173.1 | 68.0 | 749.3 | 1412.5 | 14.5 | 1.8 |
| 14 | 12.1 | 12.4 | 152.4 | 107.6 | 50.2 | 461.2 | 891.3 | 10.9 | 1.5 |

## 4. Estimating Permeability from NMR $T_2$ Distribution

The pore-geometry factor $c_s$ accounts for the relationship between pore surface area and volume in the NMR relaxation model. For idealized cylindrical and spherical pores, $c_s$ = 4 and $c_s$ = 6, respectively. However, natural carbonate rocks exhibit complex and irregular pore geometries that cannot be adequately described by either idealization. In the absence of independent pore-scale geometric measurements, we adopted an intermediate effective value of $c_s$ = 5 to estimate $k$ using the physics-based model, Eq. (5). This assumption does not imply that carbonate pores possess a uniform or intermediate geometry, but provides a practical approximation for converting NMR relaxation times into characteristic pore sizes. Although permeability predictions depend quadratically on $c_s$ in the adopted CPA formulation, varying this factor between 4 and 6 changes predicted permeability by factors of 0.64-1.44 relative to $c_s$ = 5, with all other parameters held constant. Therefore, uncertainty associated with this approximation is expected to be relatively modest compared with the potentially larger uncertainties associated with surface relaxivity and the representation of pore-network connectivity. Nevertheless, $c_s$ should be regarded as an effective parameter whose value may vary among carbonate pore systems.

Following Skaggs (2011), the CPA proportionality constant was assumed to be $C_{CPA}$ = 72.2. In the CPA permeability model, Eq. (3), the coefficient $C_{CPA}$ depends on the underlying theoretical assumptions, the definition of the critical pore-throat length scale, and pore-network characteristics. Katz and Thompson (1986) originally proposed $C_{CPA}$ = 226, whereas subsequent studies have considered alternative values. Ghanbarian et al. (2016) systematically evaluated several previously proposed coefficients for tight-gas sandstones and found that $C_{CPA}$ = 72.2 was among those yielding accurate permeability estimates, although $C_{CPA}$ = 53.5 provided the best agreement with their measurements. Subsequently, Ghanbarian (2020) applied CPA with $C_{CPA}$ = 72.2 to 105 uniform sand and glass-bead packs and obtained an RMSLE of 0.16. Based on these previous evaluations, we adopted $C_{CPA}$ = 72.2 as the baseline coefficient in the present study and further assessed its influence on permeability predictions through sensitivity analysis.

Because effective transverse surface relaxivity $\rho_{2e}$ was not independently determined for the Morro do Chaves samples, a literature-based estimate was adopted. Jácomo et al. (2025) reported effective surface relaxivity values ranging from 0.22 to 18.65 μm/s for nine carbonate samples from the Barra Velha Formation, based on the integration of NMR relaxation measurements and mercury intrusion capillary pressure (MICP) data. The arithmetic mean of these values, $\rho_{2e}$ = 7.2 μm/s, was adopted in the present study. This estimate is also reasonably close to the mean value of 8.05 μm/s reported by August et al. (2025) for four lithofacies from the same formation. However, the Morro do Chaves Formation is dominated by reworked bioclastic coquinas and is more closely comparable, in terms of lithofacies, to the bioclastic coquina intervals of the Itapema Formation than to the carbonate facies represented by the Barra Velha samples. To the best of our knowledge, no published, methodologically comparable surface relaxivity estimates are available for the Itapema Formation. Furthermore, because the literature-derived values were inferred from relationships between NMR relaxation characteristics and MICP-derived pore-throat dimensions, $\rho_{2e}$ should be interpreted as an effective conversion parameter rather than a direct measurement of intrinsic surface relaxivity. Such relationships do not necessarily establish that the inferred characteristic length scale corresponds to the critical pore-throat diameter governing permeability. Accordingly, the adopted value of $\rho_{2e}$ = 7.2 μm/s represents an externally constrained approximation whose applicability to the Morro do Chaves carbonates remains uncertain because of potential differences in mineralogy, pore structure, and pore-body-to-throat relationships. Independent measurements of NMR relaxation characteristics and critical pore-throat dimensions on the same samples would be necessary to evaluate this approximation directly.

To compare the predictive performance of the proposed physics-based model with empirical NMR permeability relationships, we evaluated the generalized TC and SDR formulations calibrated by Machado et al. (2008) for carbonate samples from the Santos Basin, Brazil, as well as other models reported in Table 1. Martins (2019) previously calibrated generalized TC and SDR models using 45 carbonate samples from the Morro do Chaves Formation. However, those models were not directly evaluated in the present study because the current dataset partially overlaps with the database used by Martins (2019). More specifically, the 14 samples analyzed here represent a subset of the larger dataset, and formation factor measurements were available only for these 14 samples.

To evaluate the predictive performance of the permeability models, two complementary statistical metrics were employed: the root mean squared logarithmic error (RMSLE) and the mean logarithmic prediction error (MLPE) defined as

$$RMSLE = \sqrt{\frac{\sum_{i=1}^{N}\left[\log_{10}(k_{est}^{i}) - \log_{10}(k_{meas}^{i})\right]^{2}}{N}} \tag{6}$$

$$MLPE = \frac{1}{N}\sum_{1}^{N}\log_{10}\left(\frac{k_{est}^{i}}{k_{meas}^{i}}\right) \tag{7}$$

where $N$ = 14 is the number of samples, and $k_{est}$ and $k_{meas}$ are the estimated and measured permeability values, respectively.

RMSLE quantifies the overall magnitude of prediction errors, whereas MLPE characterizes systematic prediction bias. Both metrics were calculated using base-10 logarithms. An MLPE value of zero indicates no systematic multiplicative bias, whereas positive and negative values indicate tendencies toward overestimation and underestimation of permeability, respectively. The combined use of RMSLE and MLPE enables assessment of both predictive accuracy and systematic bias without requiring calibration of model parameters against the investigated permeability dataset.

## 5. Results and Discussion

Fig. 2 shows the relationships between Klinkenberg-corrected permeability and $T_{2lm}$, NMR porosity, $T_{2peak}$, and the reciprocal of formation factor for the 14 carbonate samples from the Morro do Chaves Formation. As shown in Fig. 2a, permeability generally increases with increasing $T_{2lm}$ following a power-law trend with an exponent of approximately 1.61. However, the relatively low coefficient of determination ($R^2$ = 0.28) indicates substantial scatter in the data. We also observed a weak relationship between permeability and NMR porosity (Fig. 2b). Although stronger porosity–permeability correlations have been reported for some carbonate datasets (e.g., Shabani et al., 2023), such behavior is often associated with smaller datasets, more homogeneous lithologies, or carbonate systems with less complex pore architecture (Lucia, 1983). For example, Ng and Santamarina (2023) analyzed several hundred carbonate samples compiled from the literature and showed that permeability spanning six orders of magnitude was only weakly correlated with porosity because of the highly variable and heterogeneous pore connectivity characteristic of carbonate rocks.

Fig. 2c presents Klinkenberg-corrected permeability as a function of $T_{2peak}$ on a logarithmic scale. Compared with $T_{2lm}$, $T_{2peak}$ exhibited a stronger association with permeability ($R^2$ = 0.41 vs. 0.28). This result supports its use as the characteristic relaxation parameter evaluated here, but does not establish that uniquely represents the critical pore-throat diameter. Rather, the proposed correspondence between $T_{2peak}$ and the flow-controlling length scale should be regarded as a testable approximation.

Fig. 2d shows the Klinkenberg-corrected permeability versus the reciprocal of the formation factor ($1/F$). Following the theoretical expectation of CPA-based permeability models (Katz and Thompson, 1986; Skaggs, 2011), a linear relationship passing through the origin was fitted to the data. Although permeability generally increased with increasing $1/F$, the data remained scattered, further reflecting the heterogeneous and complex pore structure of the Morro do Chaves carbonate samples.

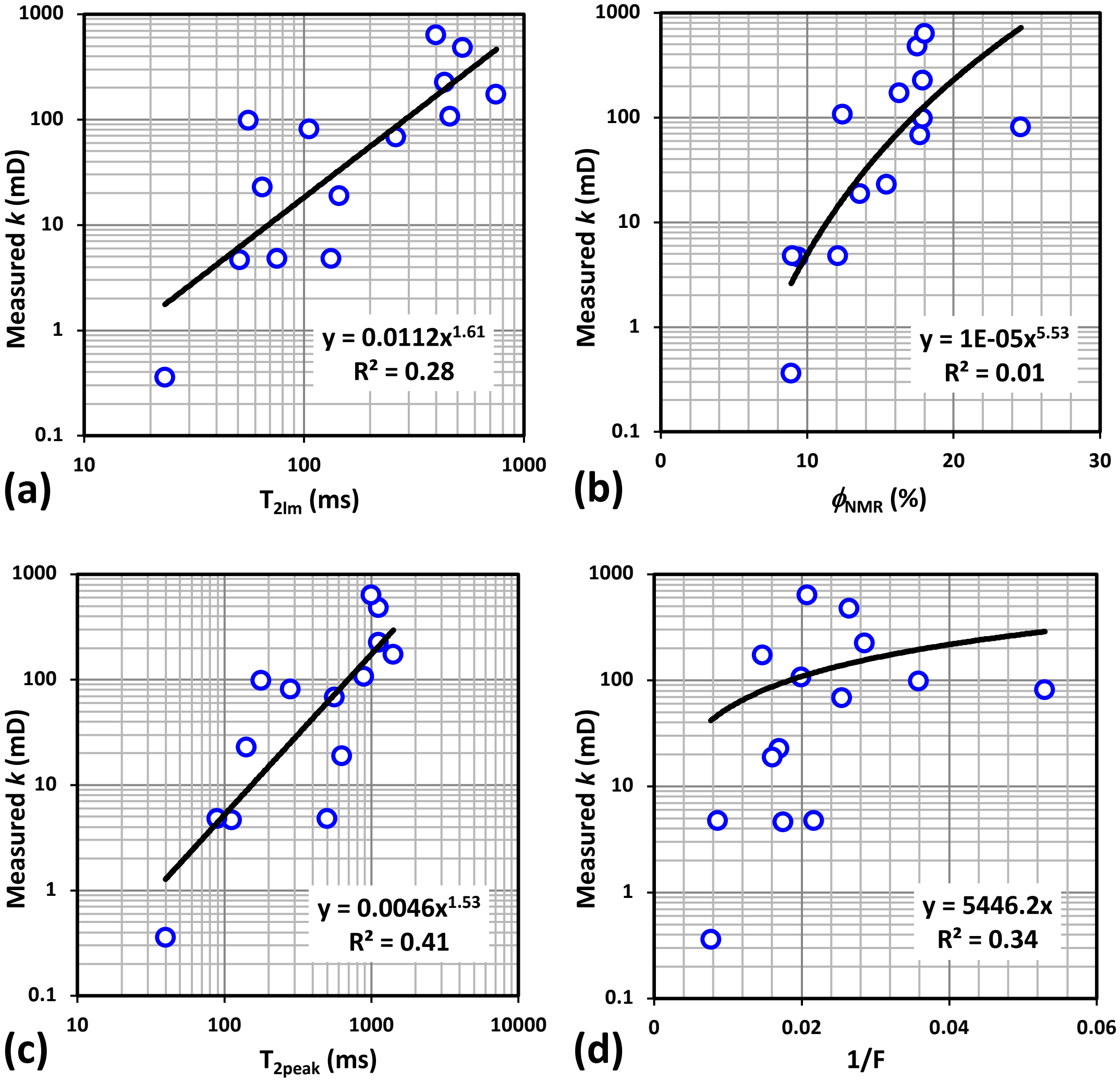

Fig. 2. Klinkenberg-corrected permeability against (a) $T_{2lm}$, (b) NMR porosity, (c) $T_{2peak}$, and (d) reciprocal of formation factor for 14 carbonate samples from the Morro do Chaves formation.

Fig. 3 compares permeability estimated using the proposed physics-based model, Eq. (5), with predictions from generalized TC and SDR models. The physics-based model produced a root mean square logarithmic error (RMSLE) of 0.56 (Fig. 3a), substantially lower than the errors associated with the generalized TC and SDR formulations (Figs. 3b and 3c). To assess the sensitivity of the CPA–NMR permeability predictions to the coefficient $C_{CPA}$, we evaluated three previously proposed values, $C_{CPA}$ = 53.5, 72.2, and 226, while maintaining the effective surface relaxivity at $\rho_{2e}$ = 7.2 μm/s and keeping all other model inputs unchanged. The corresponding RMSLE values were 0.59, 0.56, and 0.67, respectively. Thus, the baseline value of $C_{CPA}$ = 72.2 yielded the lowest prediction error among the three coefficients considered, although the differences were relatively modest. These results indicate that the overall prediction error is less sensitive to the choice of $C_{CPA}$ within the investigated range than to variations in effective surface relaxivity. Nevertheless, the appropriate coefficient may depend on the underlying CPA formulation, pore-network geometry, and definition of the characteristic critical pore-throat dimension (Skaggs, 2011). Accordingly, the present sensitivity analysis supports the use of $C_{CPA}$ = 72.2 for the investigated dataset but does not establish its universal applicability to heterogeneous carbonate rocks.

To further investigate the influence of the characteristic NMR relaxation time on permeability prediction, we evaluated Eq. (5) by replacing the dominant relaxation time, $T_{2peak}$, with the geometric mean relaxation time, $T_{2lm}$, while keeping all other model parameters unchanged. The resulting RMSLE increased from 0.56 to 0.77, corresponding to an approximately 38% increase in prediction error. Furthermore, the MLPE was -0.6 (compared to 0.1 found for $T_{2peak}$), indicating systematic underestimation of permeability when $T_{2lm}$ was used. These results suggest that $T_{2peak}$ is a more suitable characteristic relaxation time than $T_{2lm}$ for permeability estimation using the proposed CPA–NMR formulation in the investigated carbonate samples. Nevertheless, the improved predictive performance does not necessarily establish that $T_{2peak}$ directly represents the critical pore-throat diameter controlling fluid flow.

The proposed CPA–NMR formulation, Eq. (5), achieved an RMSLE of 0.56, with 13 of the 14 estimated permeability values falling within a factor of 10 of their corresponding measured values. The only exception was Sample 11, for which the estimated permeability was 97.7 mD compared with a measured value of 4.79 mD, representing an overprediction by a factor of approximately 20.6. This discrepancy highlights that, despite the relatively low overall prediction error, substantial deviations may occur for individual samples, potentially reflecting limitations in representing complex pore-throat characteristics using a single NMR-derived characteristic relaxation time.

The generalized TC model, however, systematically underestimated permeability and the generalized SDR model considerably overestimated permeability. Similar overestimation and underestimation behavior of TC- and SDR-type models has been widely reported in the literature (Dlubac et al., 2013; Yoneda et al., 2022; Gonçalves et al., 2024). For instance, Kendrick et al. (2021) reported that an SDR model calibrated for unconsolidated alluvial aquifers substantially overestimated hydraulic conductivity when applied to glacial aquifers, despite comparable hydraulic conductivity ranges among the studied formations.

The differences in predictive performance among the evaluated models were further examined using MLPE to distinguish systematic overestimation from underestimation. The proposed CPA–NMR formulation, E. (5), exhibited a relatively small positive bias (MLPE = 0.1), whereas the generalized TC and SDR formulations showed varying degrees of systematic underestimation or overestimation, depending on their calibration datasets and parameterizations. These results indicate that the proposed formulation achieved comparatively low prediction error and systematic bias for the investigated carbonate samples without local permeability calibration. Nevertheless, the limited sample size and formation-specific nature of the evaluation preclude conclusions regarding universal predictive superiority.

Recently, Chen et al. (2023) developed a data-driven approach for estimating matrix permeability in sandstone aquifers using NMR $T_2$ relaxation data and machine learning algorithms. Using a dataset of 132 sandstone core samples spanning seven orders of magnitude in permeability, they demonstrated that classical NMR permeability models such as SDR and TC produced relatively large prediction errors, whereas machine-learning approaches substantially improved predictive accuracy. In particular, the gradient boosting decision tree model achieved the best performance with a correlation coefficient of approximately 0.92. Their SHAP analysis further showed that different portions of the $T_2$ spectrum contribute differently to permeability prediction, highlighting the important role of intermediate pore sizes in controlling flow.

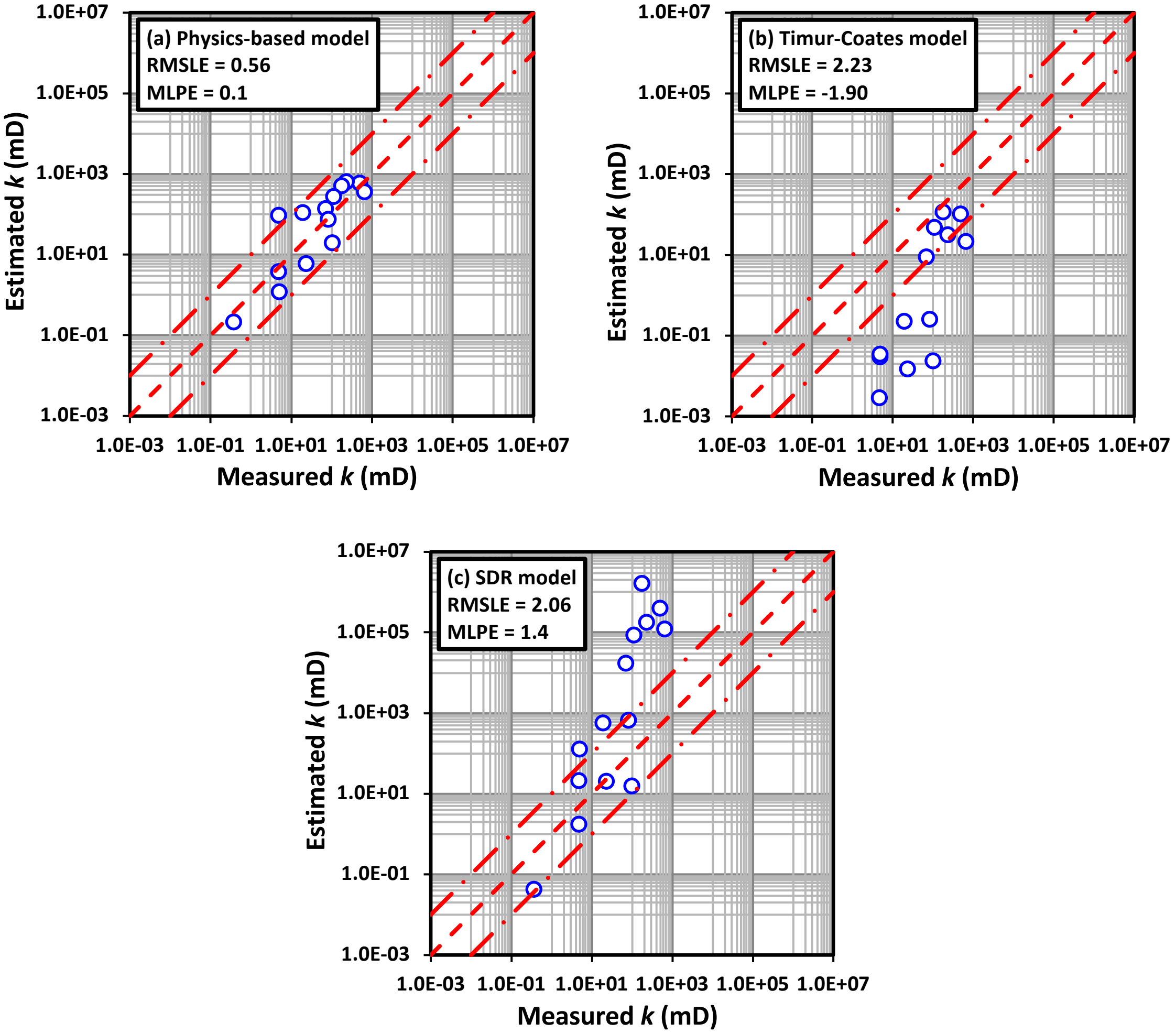

Figure 3. Estimated versus measured permeability for 14 carbonate samples analyzed in this study. Estimated permeabilities were obtained using the physics-based model (Eq. 5) as well as the generalized TC and SDR models calibrated by Machado et al. (2008).

Bias in generalized TC and SDR permeability models has been attributed to several factors, including uncertainty in selecting $T_{2cutoff}$ (Xiao et al., 2014), the presence of large but poorly connected vugs (Wang et al., 2018), organic matter (Mashhadi et al., 2025), and strong formation heterogeneity (Eeva et al., 2023). In carbonate rocks, pore connectivity may be as important as pore size itself. Wang et al.

(2018), for example, analyzed Madison limestone samples and showed that strong long-$T_2$ responses may originate from large but poorly connected vugs, causing SDR-type interpretations to overestimate permeability relative to the actual flow capacity of the rock.

An important distinction between the proposed CPA–NMR formulation, Eq. (5), and conventional SDR and TC models concerns their calibration requirements. The proposed model estimates permeability without fitting its parameters to measured permeability from the investigated carbonate samples, whereas SDR and TC formulations generally require empirical calibration to establish formation-specific coefficients. Accordingly, the present comparison evaluates the ability of a locally uncalibrated, physics-based formulation to predict permeability relative to empirical relationships calibrated using data from other formations. Although locally calibrated SDR and TC models might achieve improved predictive accuracy, such calibration requires measured permeability from the target formation and therefore represents a different application scenario. Consequently, the results should be interpreted as demonstrating the potential of the proposed approach for permeability prediction without local core-permeability calibration, rather than establishing the universal superiority of its functional form over empirical SDR and TC relationships.

The improved predictive performance of Eq. (5) relative to the generalized TC and SDR formulations may be partly attributed to the explicit incorporation of formation factor and surface relaxivity. The formation factor provides information about electrical transport through the connected pore network and is commonly related to effective electrical tortuosity through $\tau_e = \phi F$ (Ghanbarian et al., 2013), assuming negligible surface conduction. Here, $\tau_e$ represents electrical rather than geometrical or hydraulic tortuosity. Although electrical and hydraulic transport are both influenced by pore-network connectivity, tortuosity, and constrictions, their respective pathways and sensitivities to pore-throat geometry may differ, particularly in heterogeneous carbonate rocks. Thus, formation factor should be interpreted as an indirect indicator of pore-network transport characteristics rather than a direct measure of hydraulic tortuosity or connectivity. Arns et al. (2005) used three-dimensional tomographic images of Fontainebleau sandstone to evaluate permeability predictors based on pore volume-to-surface-area ratio, critical channel diameter, NMR-derived length scales, and formation factor. Their results demonstrated the importance of critical pore dimensions and formation factor in permeability estimation, highlighting the complementary roles of pore-throat geometry and electrical transport properties. Accordingly, incorporating formation factor alongside NMR-derived characteristic pore dimensions provides a physically motivated means of accounting for aspects of pore-network structure not explicitly represented in conventional TC and SDR formulations, although it does not fully resolve the differences between electrical and hydraulic transport pathways.

Table 3 summarizes the RMSLE and MLPE calculated for the permeability models listed in Table 1, excluding the formulations of Martins (2019), whose calibration dataset partially overlaps with the samples investigated in this study. Among the evaluated models, the proposed physics-based CPA–NMR formulation, Eq. (5), achieved the lowest RMSLE (0.56) and the smallest MLPE (0.1), indicating the highest predictive accuracy and lowest systematic logarithmic bias for the investigated samples. The positive MLPE indicates a slight overall tendency toward permeability overestimation.

The second-lowest RMSLE (0.78) was obtained using the generalized Timur–Coates (TC) formulation proposed by Amabeoku et al. (2001), which was calibrated using seven carbonate samples from an Arab reservoir. However, its MLPE of −0.6 indicates systematic underestimation of permeability. The generalized SDR formulations developed by Aghda et al. (2018) and Shabani et al. (2021) yielded comparable RMSLE values of 1.01 and 1.04, respectively, with corresponding MLPE values of −0.9 and −1.0, indicating systematic underestimation. Both formulations were developed using carbonate samples from the Asmari Formation in Iran. Notably, Shabani et al. (2021) reportedly calibrated their generalized SDR model using a single carbonate sample. Estimating three independent fitting coefficients from one sample would constitute an underdetermined calibration problem unless additional constraints were imposed, raising concerns regarding parameter identifiability and generalizability. Overall, these results indicate that the proposed CPA–NMR formulation, Eq. (5), provides more accurate permeability predictions than the evaluated externally calibrated empirical relationships for the present dataset, without requiring calibration against measured permeability from the investigated formation.

The results presented in Table 3 demonstrate that some empirically calibrated TC and SDR formulations can provide reasonable permeability estimates for carbonate rocks. However, empirical data-driven models are inherently database dependent, meaning that their predictive capability is often restricted to formations similar to those used during calibration (Schaap and Leij, 1998; Ghanbarian and Pachepsky, 2022). In contrast, physics-based formulations, provided that their input parameters are reasonably estimated, offer the potential to predict permeability from physically meaningful transport relationships rather than purely empirical fitting.

Table 3. Root mean squared logarithmic error (RMSLE) and mean logarithmic prediction error (MLPE) for the proposed CPA–NMR formulation, Eq. (5), and previously published SDR and Timur-Coates permeability models evaluated using 14 carbonate samples from the Morro do Chaves Formation. Both metrics were calculated using base-10 logarithms. The models of Martins (2019) were excluded from the quantitative comparison because their calibration dataset overlaps with the samples investigated in this study.

| Reference | Model | RMSLE | MLPE |
|---|---|---|---|
| Amabeoku et al. (2001) | $k = 0.0574\phi^{3.79}T_{2lm}^{1.89}$ | 1.91 | -1.9 |
| Amabeoku et al. (2001) | $k = 511.4\phi^{2.84}\left(\frac{FFI}{BVI}\right)^2$ | 0.77 | -0.6 |
| Hassall et al. (2004) | $k = 0.5\phi^2(\rho_2 T_{2lm})^2$ | 2.59 | 2.5 |
| Machado et al. (2008) | $k = 7.3 \times 10^{-11}\phi^{2.68}T_{2lm}^{4.56}$ | 2.06 | 1.4 |
| Machado et al. (2008) | $k = \left(\frac{\phi}{109.64}\right)^{1.74}\left(\frac{FFI}{BVI}\right)^{3.85}$ | 2.23 | -1.9 |
| Aghda et al. (2018) | $k = 0.0216\phi^{1.64}T_{2lm}^{1.68}$ | 1.01 | -0.9 |
| Aghda et al. (2018) | $k = \left(\frac{\phi}{0.222}\right)^{3.9}\left(\frac{FFI}{BVI}\right)^{0.51}$ | 2.18 | -2.1 |
| Martins (2019) | $k = 0.12 \times 10^4\phi^{4.48}T_{2lm}^{1.0}$ | - | - |
| Martins (2019) | $k = 3.45 \times 10^4\phi^{5.05}\left(\frac{FFI}{BVI}\right)^{0.78}$ | - | - |
| Shabani et al. (2021) | $k = 0.0004\left(\frac{\phi}{10}\right)^3 T_{2lm}^{1.6}$ | 1.04 | -1.0 |
| Shabani et al. (2021) | $k = 0.022\left(\frac{\phi}{10}\right)^4\left(\frac{FFI}{BVI}\right)^2$ | 2.04 | -2.0 |
| This study | Eq. (5) | 0.56 | 0.1 |

An important limitation of the present study is the relatively small number of investigated samples and their origin from a single geological formation. The objective was not to establish the statistical representativeness of these samples with respect to the Morro do Chaves Formation, but rather to assess whether the CPA–NMR formulation, Eq. (5), could estimate permeability with reasonable accuracy without calibration against locally measured permeability, given the availability of NMR $T_2$ relaxation distributions and independently measured electrical formation factors. Although the results support the feasibility of this approach for the investigated samples, the limited sample size introduces uncertainty into the estimated prediction errors, and the observed performance does not establish applicability to other carbonate formations or lithologies. Broader validation is constrained by the limited availability of datasets containing these complementary measurements for the same samples. Furthermore, the effective surface relaxivity adopted from a different carbonate formation may not adequately represent rocks with substantially different mineralogical compositions or pore structures. Consequently, the present findings should be interpreted as an initial evaluation of the proposed formulation, and its predictive reliability and transferability should be investigated using larger, independent datasets encompassing a broader range of geological formations and pore-system characteristics.

## 6. Sensitivity to the surface relaxivity

To evaluate the sensitivity of the proposed permeability model to the assumed effective transverse surface relaxivity, $\rho_{2e}$, we performed a scenario-based sensitivity analysis using values of 0.22, 2.0, 7.2, 12.0, and 18.65 μm/s. The lower and upper bounds correspond to the range reported by Jácomo et al. (2025) for carbonate samples from the Barra Velha Formation, whereas 7.2 μm/s represents the baseline value adopted in this study. For each scenario, permeability was estimated for all 14 carbonate samples using Eq. (5), while maintaining the remaining model parameters unchanged. The corresponding RMSLE was then calculated to quantify the influence of surface relaxivity on prediction accuracy. Because permeability scales quadratically with $\rho_{2e}$, this analysis provides a direct assessment of the sensitivity of the model predictions to uncertainties associated with transferring surface relaxivity estimates between carbonate formations.

Results presented in Table 4 demonstrate that permeability predictions are highly sensitive to the assumed surface relaxivity. At the lower bound of $\rho_{2e}$ = 0.22) μm/s, the model systematically underestimated permeability, with an MLPE of −2.9 and an RMSLE of 2.98. Increasing $\rho_{2e}$ to 2.0 μm/s substantially reduced the RMSLE to 1.15, although permeability remained predominantly underestimated. The lowest RMSLE among the five scenarios (0.56) was obtained using the baseline value of 7.2 μm/s, for which the MLPE was approximately 0.1, indicating relatively small overall systematic bias. Further increases in $\rho_{2e}$ to 12.0 and 18.65 μm/s increased the RMSLE to 0.77 and 1.08, respectively, accompanied by increasingly positive MLPEs of 0.5 and 0.9. These results indicate a transition from systematic underestimation at low surface relaxivity to systematic overestimation at higher values.

The pronounced sensitivity of permeability predictions to $\rho_{2e}$ is consistent with the quadratic dependence in Eq. (5). Relative to the baseline value of 7.2 μm/s, adopting the lower and upper bounds changes the predicted permeability by multiplicative factors of approximately 0.00093 and 6.71, respectively. Nevertheless, the baseline value produced the lowest RMSLE among the five scenarios, suggesting that the literature-based estimate of 7.2 μm/s provides a reasonable effective surface relaxivity for the investigated samples. This agreement should not, however, be interpreted as independent validation of the assumed value or evidence that surface relaxivity is uniform across the samples. The sensitivity analysis assumes a common $\rho_{2e}$ for all 14 samples, whereas variations in mineralogy, pore geometry, and surface chemistry may produce sample-specific differences in effective surface relaxivity. Furthermore, the literature-derived range represents variability observed in the Barra Velha Formation rather than a statistically defined uncertainty interval for the Morro do Chaves samples. Consequently, although the proposed model exhibits its lowest prediction error near the adopted baseline value among the scenarios evaluated, its predictive accuracy and transferability remain dependent on reliable characterization of surface relaxivity. Independent measurements of $\rho_{2e}$ on the investigated samples and additional carbonate formations are therefore needed to establish the broader applicability of the proposed framework.

Table 4. Sensitivity of permeability estimates from Eq. (5) to effective transverse surface relaxivity $\rho_{2e}$ for 14 carbonate samples from the Morro do Chaves Formation. Estimated permeability $k_{est}$, 10-based logarithmic residuals $\log(k_{est}/k_{meas})$, mean logarithmic residuals, and RMSLE are reported for five assumed $\rho_{2e}$ values. The baseline value of 7.2 μm/s and the lower and upper literature bounds of 0.22 and 18.65 μm/s are based on Jácomo et al. (2025).

| **Sample** | **$\rho_{2e}$ = 0.22 μm/s** | | **$\rho_{2e}$ = 2 μm/s** | | **$\rho_{2e}$ = 7.2 μm/s** | | **$\rho_{2e}$ = 12 μm/s** | | **$\rho_{2e}$ = 18.65 μm/s** | |
|---|---|---|---|---|---|---|---|---|---|---|
| | **$k_{est}$ (mD)** | **$log(\frac{k_{est}}{k_{meas}})$** | **$k_{est}$ (mD)** | **$log(\frac{k_{est}}{k_{meas}})$** | **$k_{est}$ (mD)** | **$log(\frac{k_{est}}{k_{meas}})$** | **$k_{est}$ (mD)** | **$log(\frac{k_{est}}{k_{meas}})$** | **$k_{est}$ (mD)** | **$log(\frac{k_{est}}{k_{meas}})$** |
| 1 | 0.1347 | -2.7 | 11.13 | -0.8 | 144.3 | 0.3 | 400.8 | 0.8 | 968.1 | 1.2 |
| 2 | 0.0057 | -3.6 | 0.47 | -1.7 | 6.1 | -0.6 | 16.9 | -0.1 | 40.9 | 0.3 |
| 3 | 0.5580 | -2.9 | 46.12 | -1.0 | 597.7 | 0.1 | 1660.3 | 0.5 | 4010.4 | 0.9 |
| 4 | 0.0037 | -3.1 | 0.30 | -1.2 | 4.0 | -0.1 | 11.0 | 0.4 | 26.5 | 0.8 |
| 5 | 0.0012 | -3.6 | 0.10 | -1.7 | 1.2 | -0.6 | 3.4 | -0.1 | 8.3 | 0.2 |
| 6 | 0.6012 | -2.6 | 49.69 | -0.7 | 643.9 | 0.5 | 1788.7 | 0.9 | 4320.6 | 1.3 |
| 7 | 0.0190 | -3.7 | 1.57 | -1.8 | 20.3 | -0.7 | 56.5 | -0.2 | 136.5 | 0.1 |
| 8 | 0.3482 | -3.3 | 28.77 | -1.3 | 372.9 | -0.2 | 1035.8 | 0.2 | 2502.0 | 0.6 |
| 9 | 0.0705 | -3.1 | 5.83 | -1.1 | 75.5 | 0.0 | 209.8 | 0.4 | 506.6 | 0.8 |
| 10 | 0.0002 | -3.2 | 0.02 | -1.3 | 0.2 | -0.2 | 0.6 | 0.2 | 1.5 | 0.6 |
| 11 | 0.0913 | -1.7 | 7.54 | 0.2 | 97.8 | 1.3 | 271.6 | 1.8 | 656.0 | 2.1 |
| 12 | 0.1071 | -2.2 | 8.85 | -0.3 | 114.8 | 0.8 | 318.8 | 1.2 | 770.0 | 1.6 |
| 13 | 0.4917 | -2.5 | 40.63 | -0.6 | 526.6 | 0.5 | 1462.8 | 0.9 | 3533.3 | 1.3 |
| 14 | 0.2652 | -2.6 | 21.91 | -0.7 | 284.0 | 0.4 | 788.9 | 0.9 | 1905.5 | 1.2 |
| | **MLPE** | **-2.9** | | **-1.0** | | **0.1** | | **0.5** | | **0.9** |
| | **RMSLE** | **2.98** | | **1.15** | | **0.56** | | **0.77** | | **1.08** |

## 7. Conclusions

In this study, we developed and evaluated a physics-based permeability estimation approach that integrates NMR transverse relaxation characteristics, electrical formation factor, and effective surface relaxivity within a critical path analysis (CPA) framework. The proposed CPA–NMR formulation, Eq. (5), was evaluated using 14 heterogeneous carbonate samples from the Morro do Chaves Formation without calibrating model parameters against their measured permeability. The model achieved a root mean squared logarithmic error (RMSLE) of 0.56, with 13 of the 14 permeability predictions falling within one order of magnitude of the measured values. Its predictive accuracy exceeded that of the externally calibrated generalized Timur–Coates (TC) and Schlumberger–Doll Research (SDR) formulations considered in this study. However, this comparison reflects the performance of the selected models under conditions without local permeability calibration and does not establish the inherent superiority of the CPA–NMR formulation over locally calibrated empirical alternatives.

The results suggest that combining NMR-derived characteristic length scales with independently measured electrical formation factor provides a physically motivated approach to permeability estimation in heterogeneous carbonate rocks. Within the proposed formulation, the dominant relaxation time, $T_{2peak}$, yielded more accurate predictions than the geometric mean relaxation time, $T_{2lm}$. Nevertheless, the improved predictive performance does not independently establish that $T_{2peak}$ represents the critical pore-throat dimension controlling fluid flow. Similarly, although formation factor provides complementary information about electrical transport through the connected pore network, the relative contributions of pore geometry, electrical connectivity, and effective surface relaxivity to the observed predictive improvement remain unresolved. Sensitivity analysis further demonstrated a strong dependence of predicted permeability on effective surface relaxivity, with RMSLE increasing from 0.56 at the baseline value of 7.2 μm/s to 2.98 and 1.08 at the lower and upper literature bounds, respectively.

The principal limitations of this study are the relatively small dataset, the absence of independently measured effective surface relaxivity and critical pore-throat dimensions, and the uncertainty associated with applying externally determined model parameters to the investigated carbonate samples. One sample exhibited a permeability overprediction of approximately 20-fold, illustrating the potential limitations of representing complex pore systems using a single characteristic NMR relaxation time. Consequently, the present findings constitute an initial evaluation of the CPA–NMR formulation rather than evidence of its general transferability across geological formations or lithologies. Future studies should evaluate the approach using larger, independent datasets containing NMR relaxation distributions, electrical formation factor, measured permeability, and independently constrained effective surface relaxivity for the same samples. Complementary pore-network characterization, including mercury intrusion capillary pressure measurements or three-dimensional imaging, would also help establish the relationship between NMR-derived characteristic dimensions and the critical pore-throat constrictions governing permeability.

**Acknowledgement**

BG is grateful to the University of Texas at Arlington for financial support through faculty startup fund and the STARs award. The authors used AI-assisted language editing tools to improve grammar and readability. All scientific interpretations and conclusions remain the responsibility of the authors.

**Conflict of Interest**

The authors declare no conflicts of interest relevant to this study.

**Data Availability Statement**

Data are available and are given by Ghanbarian et al. (2026).